\documentclass[3p,twocolumn]{elsarticle} 
\usepackage[latin1]{inputenc}         

\usepackage{amsmath}
\usepackage{units}

\journal{Nucl. Inst. Meth. A (submitted to proceedings HSTD7)}

\begin{document}

\begin{frontmatter}

\author[fm]{Frank Meier}
\ead{frank.meier@psi.ch}
\author{on behalf of the Tracker Alignment group of the CMS collaboration} 
\address[fm]{Paul~Scherrer~Institut, 5232~Villigen, Switzerland and ETH~Zurich, Institute~for~Particle~Physics (IPP), Schafmattstrasse~20, 8093~Z\"urich, Switzerland}
\title{First Alignment of the Complete CMS Tracker}
\date{}
\begin{abstract}
This conference proceeding presents the first results of the full CMS Tracker alignment based on several million reconstructed tracks from the cosmic data taken during the commissioning runs with the detector in its final position and magnetic field present. The all-silicon design of the CMS Tracker poses new challenges in aligning a complex system with 15\,148~silicon strip and 1440~silicon pixel modules. For optimal track-parameter resolution, the position and orientation of its modules need to be determined with a precision of about one micrometer. The modules, well illuminated by cosmic ray particles, were aligned using two track-based alignment algorithms in sequence in combination with survey measurements. The resolution in all five track parameters is controlled with data-driven validation of the track parameter measurements near the interaction region, and tested against prediction with detailed detector simulation. An outlook for the expected tracking performance with the first proton collisions is given.
\end{abstract}
\end{frontmatter}





\graphicspath{{./img/}}


\newcommand{\vekt}[1]{\mathbf{#1}}
\newcommand{\uklamm}[2]{\underset{#2}{\underbrace{#1}}}


\newcommand{\degree}[0]{$^\circ$}


\section{Introduction}

Silicon tracking detectors in general purpose detectors like the Compact Muon Solenoid (CMS) at CERN \cite{cmsjinst} are built to reconstruct charged particles trajectories (tracks). In a magnetic field, they are described by a helix. The track parameters are the curvature $1/p_T$ (expressed as inverse transverse momentum), the impact parameters $d_{xy}$ and $d_{z}$ in the $xy$ plane and along the principal axis of the experiment respectively and the polar angles $\theta$ and $\phi$.\footnote{The CMS coordinate system is defined as follows\cite{craft08}: The origin is at the nominal collision point, the $x$-axis pointing to the center of the LHC, the $y$-axis pointing up and the $z$-axis along the anticlockwise beam direction. $\theta$ is measured from the positive $z$-axis and $\phi$ from the positive $x$ axis. The radius $r$ denotes the distance from the $z$-axis.} Their precise and accurate determination are paramount for the operation of tracking detectors with spatial resolution of the order of $\unit[10]{\mu m}$. Therefore the position of the modules needs to be known to better than this precision, which can be achieved by high mounting precision, stable frame, survey measurements and track based alignment. This article describes the track based alignment of the CMS inner tracker and the results obtained using cosmic ray particles. Brief statements will be made on the use of survey information.

\subsection{The alignment problem}

Track based alignment can be described as a \emph{least squares minimization} problem where the data from hits generated by tracks are used. A single residual $\vekt{r}_{ij}$ for hit $i$ along track $j$ is the three dimensional distance between the predicted hit location from the track model and the physical hit information from the modules, calculated using the current knowledge of the geometry. Together with the covariance matrix $\vekt{V}$ the expression to be minimized is given in equation \eqref{eqn.chi2ali}:
\begin{equation} \label{eqn.chi2ali} \chi^2(\vekt{p},\vekt{q}) 
	= \sum_j^\text{tracks} \sum_i^\text{hits} \vekt{r}_{ij}^T(\vekt{p},\vekt{q}_j) \,\vekt{V}_{ij}^{-1}\, \vekt{r}_{ij}(\vekt{p},\vekt{q}_j)
\end{equation}
where $\vekt{p}$ denotes the alignment parameters describing the current geometry and $\vekt{q}_j$ denotes the track parameters of the $j^\text{th}$ track. In principle, this can be solved using standard techniques like solving normal equations.

\begin{figure*}
	\begin{center}
	\includegraphics[width=13cm]{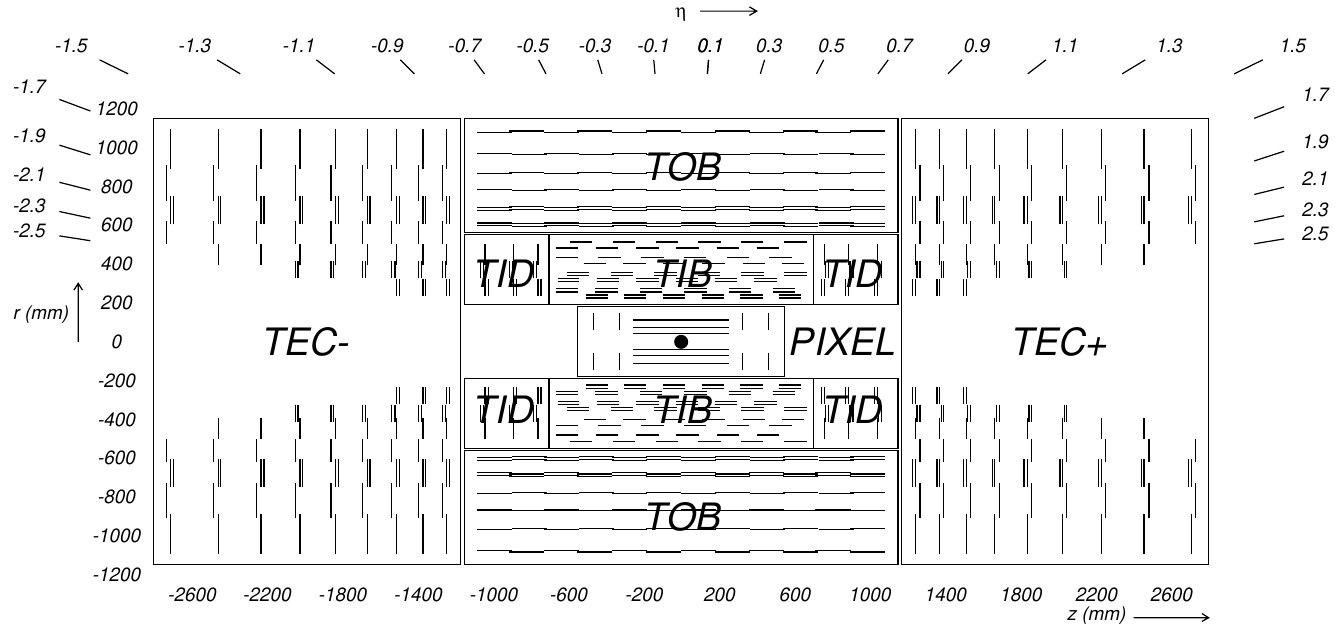}
	\end{center}
	\caption[Schematic view of the CMS inner tracker]{\textbf{Schematic view of the CMS inner tracker.} The tracker consists of several subdetectors. The innermost part is the pixel detector (a barrel and two endcaps at each side) surrounded by two barrel shaped strip detectors (TIB: tracker inner barrel, TOB: tracker outer barrel) and the endcap structures (TID: tracker inner disks, TEC: tracker endcap). \cite{cmsjinst}
}
	\label{fig.tracker}
\end{figure*}

The inner tracker at CMS consists of 1440~silicon pixel modules and 15\,148~silicon strip modules (figure \ref{fig.tracker}). Each module has six degrees of freedom, described in local coordinates $u,v,w$ with respect to the geometric center of the module and rotations $\alpha, \beta, \gamma$ around these axes. In total we have to determine 99\,528~parameters. For a typical alignment of the CMS inner tracker, around $10^6$ to $10^7$ tracks are required, depending on which hierarchy levels (modules or larger units) are selected as objects to be aligned. Therefore the number of parameters to be determined in this procedure becomes at least of the order $O(10^7)$. Solving it within hours, as required for prompt alignment, is beyond the limit of the capabilities of the computers available to the experiment.

\section{Alignment algorithms used}

Two alignment algorithms were used to produce the results reported later in this article. Both aim to reduce the complexity of the problem so that it can be solved within hours on standard CPU's\footnote{All calculations were carried out on a batch farm at CERN consisting of nodes having 2~KSi2k on average (KSi2k: Standard Performance Evaluation Corporation benchmark of Kilo Specmarks Integer year 2000, \texttt{http://www.spec.org/cpu2000/}). Job parallelization among computer nodes is used whenever suitable and reasonable.}. They are distinguished by their scope:
\subsection{Global algorithm}
Solving the full alignment problem would produce estimates for the alignment parameters and the track parameters. In our case, only the first are of interest. Restricting the solution to the alignment parameters reduces the complexity to $O(10^5)$ in our case. Using a clever scheme for setting up the matrix of the normal equations, this can be achieved using block matrix operations while using the full information from the track parameters. This is implemented in \emph{Millepede-II}\cite{blobel06}, an algorithm widely used for alignment purposes. Its advantages are that it takes all correlations between modules and higher hierarchies into account. The algorithm works in a single step, however due to rejection of tracks with too high $\chi^2$ (outlier rejection), a small number of iterations are still required. The implementation in the CMS software framework uses a simplified helix model, in which material effects due to $dE/dx$ are taken into account, but multiple scattering is currently ignored. This is a major disadvantage, as it limits the maximum resolution obtainable. At the time of this study, a memory limit allowed for an alignment of 46\,340~parameters at maximum in one step. An incremental procedure to align parts of the detector was used to overcome this limitation. To give an idea of the performance, typical time consumption (start-to-end) for a full alignment was about 4~hours.
\subsection{Local algorithm}
By assuming no track parameter dependence -- dropping correlations between alignment parameters between modules -- the problem can be reduced to solving the equation for single modules. Correlations between modules are recovered by iteration. The residuals $\vekt{r}_{ij}$ are calculated as the distance between the physical hit data and the impact point from the track using the reconstruction procedure without the hit in consideration. This is implemented in the \emph{HIP-algorithm}\cite{CMSnote-IN2006018}. The major advantage of the implementation is the use of the same track model as in the track reconstruction (K\'alm\'an filter) and therefore all material effects are taken into account. On the other hand this algorithm convergences very slowly when the start geometry is not sufficiently close to reality. Typical time consumption for a full alignment was about 5~hours.

\subsection{Combined operation}
Both algorithms make use of job parallelization on the computer cluster for data collection steps (typically up to 100~computer nodes, no intercommunication among concurrent jobs), while final calculations are carried out on a single machine. As the approaches are complementary, we used a combined method to benefit from the strength of both algorithms and to overcome their weaknesses. 
\begin{enumerate}
\item The global algorithm started from design geometry. This resolved global movements and ended up in a geometry close enough to reality for efficient operation of the local algorithm. Despite the fact that the global algorithm is capable of aligning on several hierarchical levels simultaneously, the already mentioned parameter limitation required the splitting into several steps. 
\item The local algorithm started from the outcome of the global one and resulted in a refined geometry.
\end{enumerate}
Some of the plots in the result section will show the outcome of the individual algorithms together with the combined approach. A detailed description of all steps involved can be found in \cite{craft08}.

\subsection{Survey information}
Survey data may come from optical surveys and coordinate measuring machines and are usually collected prior to or during installation. Alignment constants from such operations can be used as 
\begin{enumerate}
\item starting points for the alignment. This may enhance the convergence of an alignment algorithm, but the survey information looses its weight after the very first iteration.
\item additional data for the alignment algorithm. In the local algorithm, this can easily be done by extending the sum of equation \eqref{eqn.chi2ali}. The residuals for that are calculated as the difference between the position from the survey and the current reference geometry.
\end{enumerate}
Only the local algorithm used survey information in the results presented here.

\section{Results from commissioning with cosmic rays}

The results presented here are based on data collected in autumn 2008 during a period of cosmic ray data taking with a magnetic field of \unit[3.8]{T} in the tracker volume. The total number of events detected by CMS during this campaign was about 300~million, of which 3.2~million have hits in the tracker suitable for alignment. A cut on $p_T > \unit[4]{GeV/c}$ has been applied. The rate was about \unit[5]{Hz}. The fraction of tracks crossing the pixel detector was 3\% in the barrel and 1.5\% in the endcaps. Data used for alignment and validation were not statistically independent due to the limited number of events collected. Several low- and high-level approaches have been used to estimate and validate the alignment performance.

CMS is designed primarily for tracks originating from the nominal intersection point (including tracks from displaced vertices) and not for cosmic rays. For the tracker, this means that alignment is limited to parts with sufficient illumination from cosmic particles. We are also prone to deformation modes of the tracker which leaves the $\chi^2$ invariant. A known case is an elongation of the tracker along the $z$-axis, which is difficult to align using cosmic tracks only.

\subsection{Track $\chi^2$ distribution}

For each track of a data sample, the track $\chi^2$ is calculated. This is merely the second sum in equation \eqref{eqn.chi2ali}, weighted by the number of degrees of freedom (ndof).
\begin{equation} \label{eqn.chi2track} \frac{\chi^2_\text{track}}{\text{ndof}} 
	= \frac{1}{\text{ndof}} \sum^\text{hits}_{i} \vekt{r}_{i}^T(\vekt{p},\vekt{q}) \,\vekt{V}^{-1}\, \vekt{r}_{i}(\vekt{p},\vekt{q})
\end{equation}
A histogram of the distribution of these $\chi^2_\text{track}$ allows for a low-level evaluation of the alignment. 
\begin{figure}
	\begin{center}
	\includegraphics[width=6.1cm]{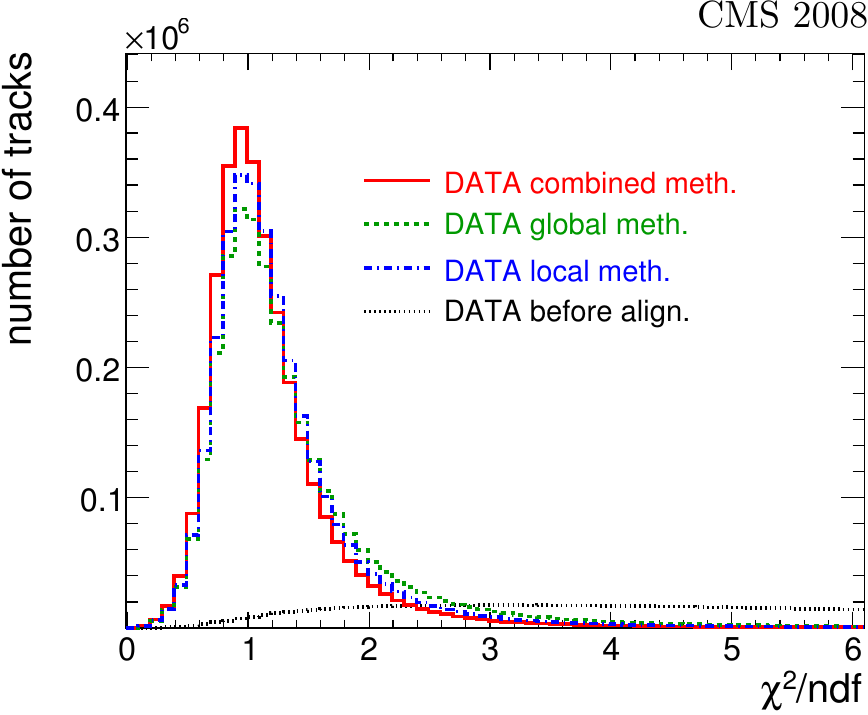}
	\end{center}
	\caption{\textbf{Track $\chi^2/\text{ndof}$ distribution.} This plot shows the distributions for the unaligned (dotted) tracker and after aligning using the combined approach (solid). For comparison, the results after the alignment using the local (dashed-dotted) and global methods (dashed) are included.}
	\label{fig.trackchi2}
\end{figure}
The results are shown in figure \ref{fig.trackchi2}, where the improvement from the unaligned to the aligned detector is clearly visible. The combined approach shows the best alignment performance.

\subsection{Distribution of the mean of the residuals (DMR)}

\begin{figure*}
	\begin{center}
	\includegraphics[width=6cm]{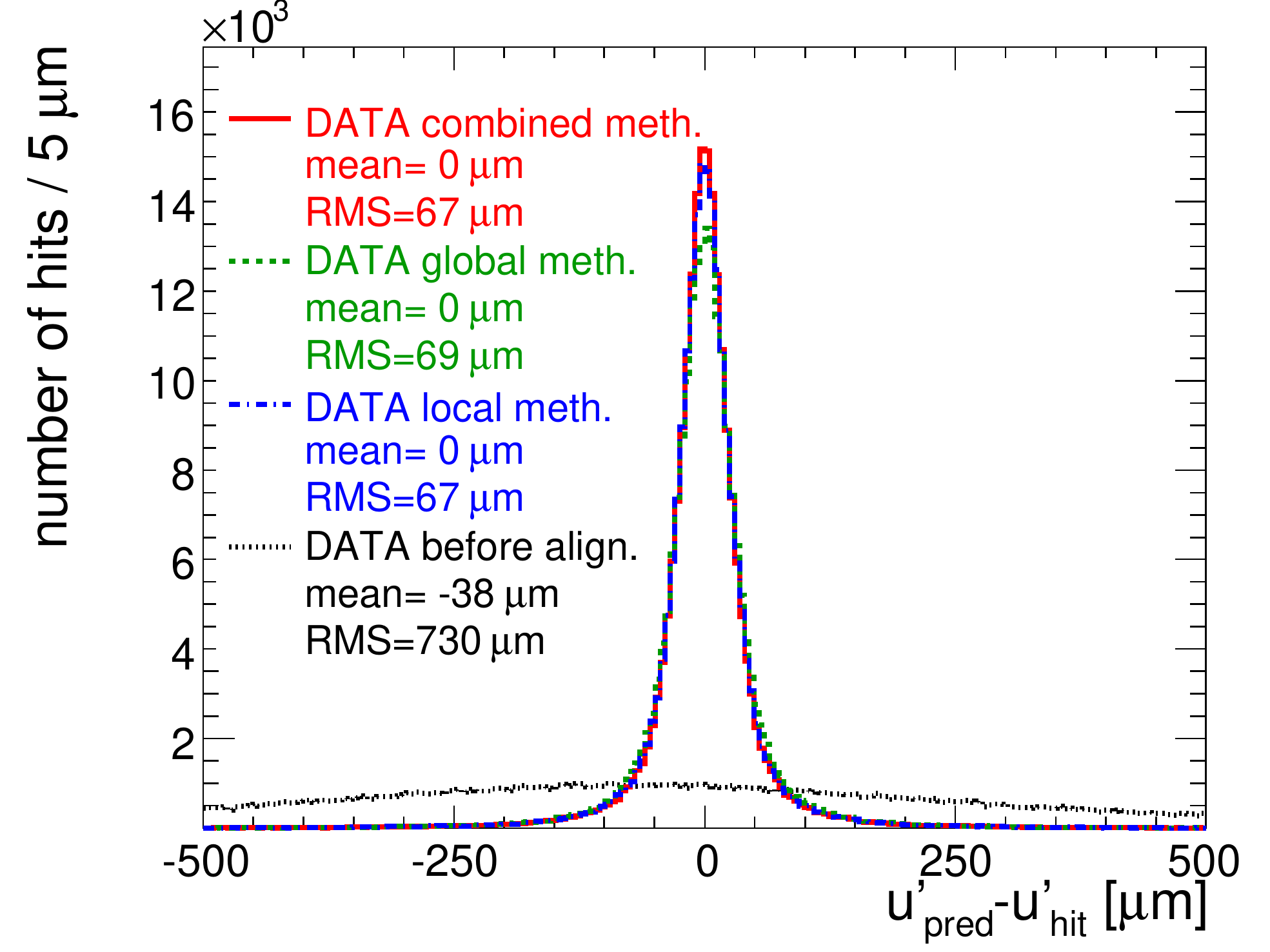}
	\includegraphics[width=6cm]{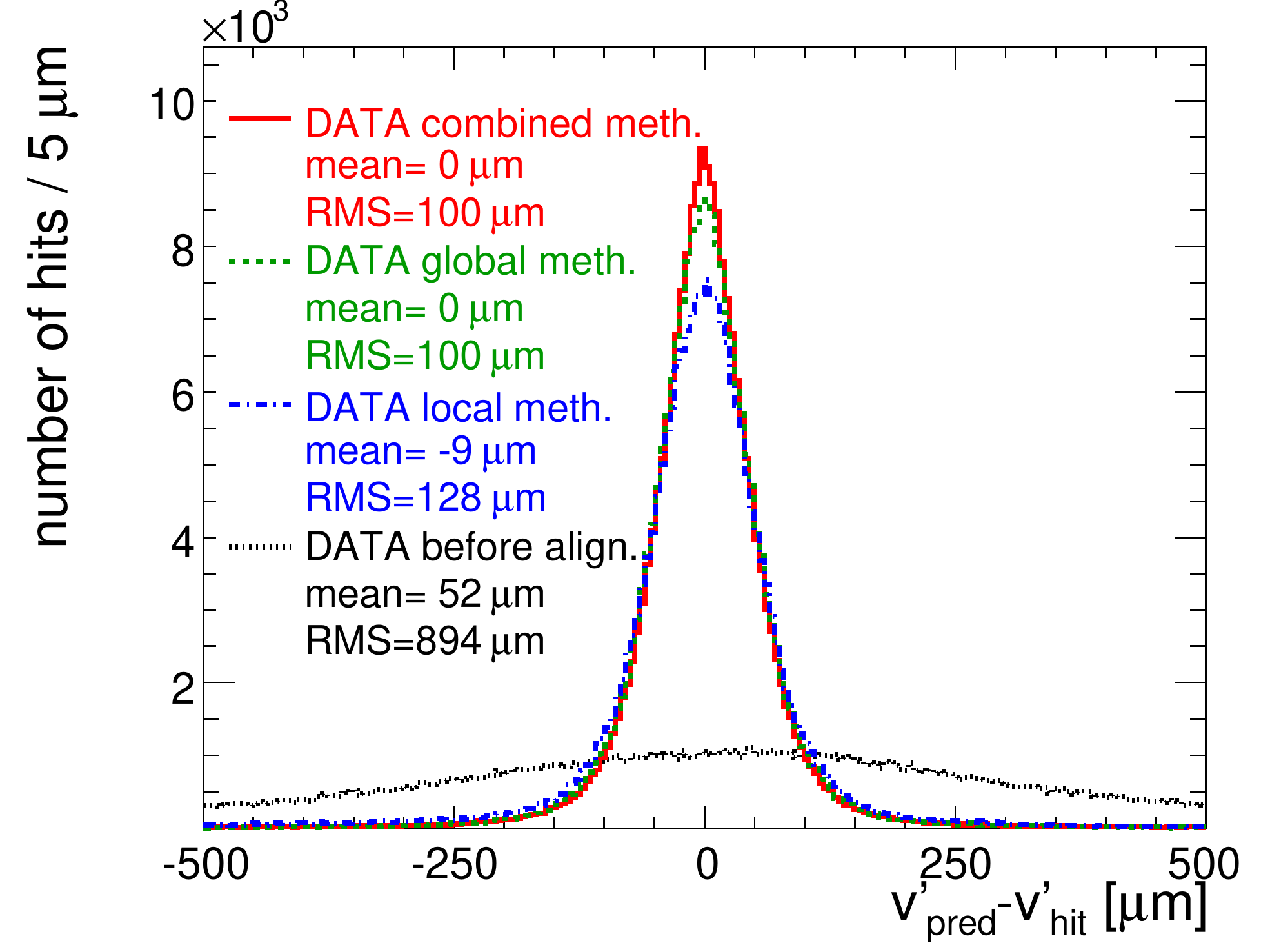}
	\includegraphics[width=6cm]{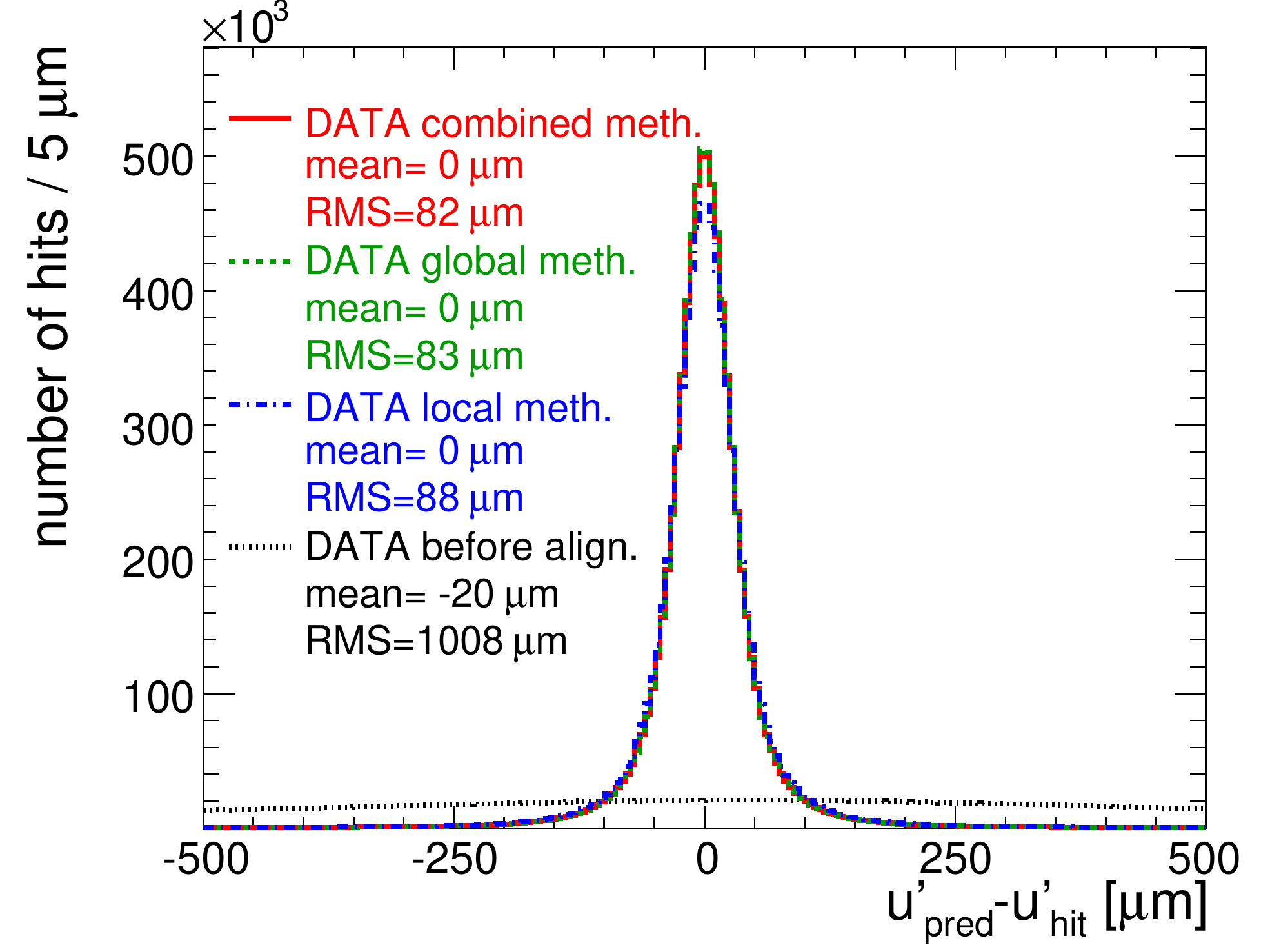}
	\includegraphics[width=6cm]{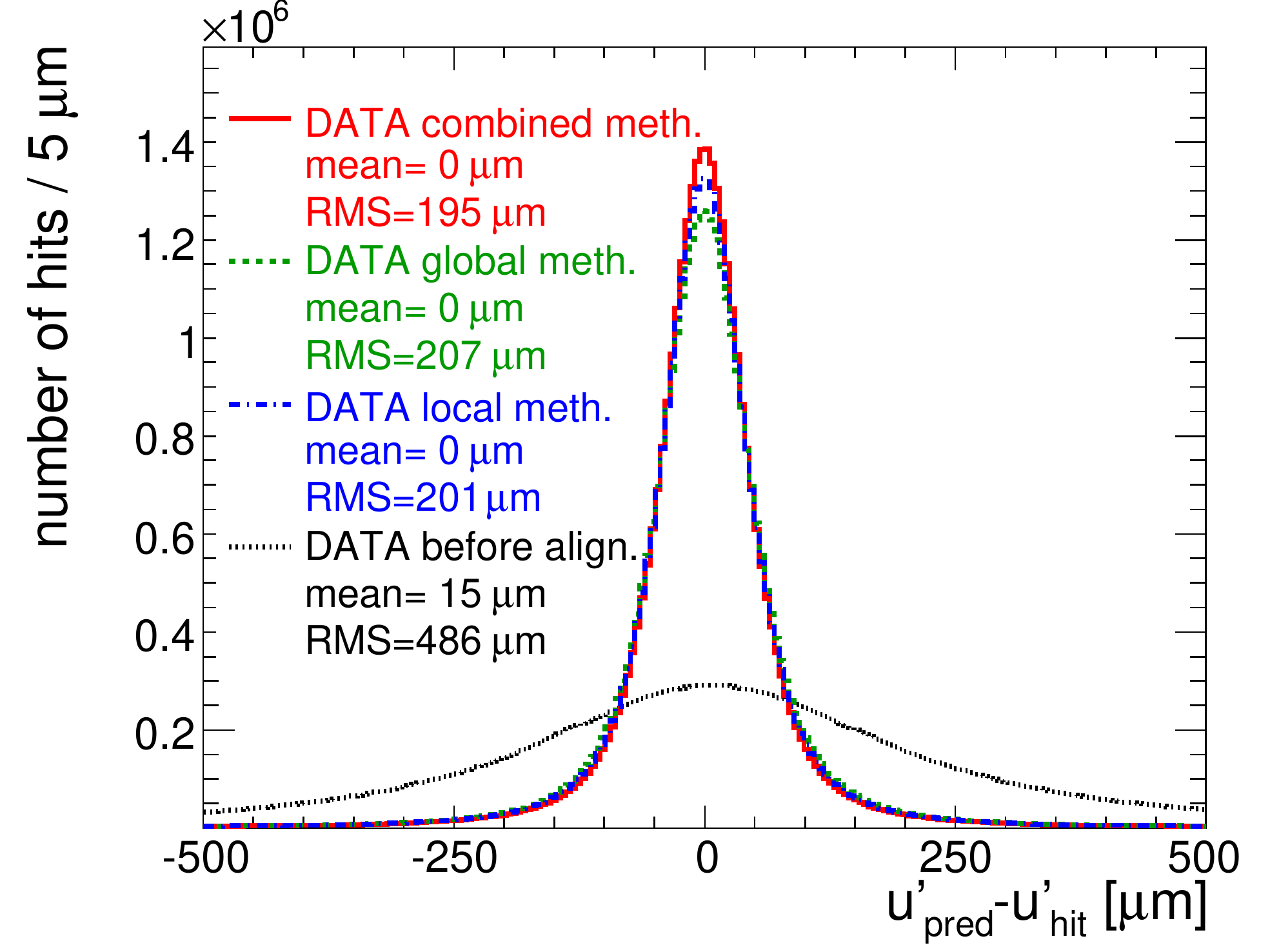}
	\end{center}
	\caption{\textbf{Some selected plots of the DMR.} The upper two plots show the distributions in the pixel barrel for the local $u$ and $v$ coordinates. Below, the plots for the TIB and TOB are shown. More details are given in table \ref{tab.dmrresults}.}
	\label{fig.dmrplots}
\end{figure*}

For each hit in a track of a data sample, the residual is calculated between the predicted position from the track and the actual hit where the hit has been removed from the track reconstruction in order to be unbiased by the hit under consideration. Such distributions were obtained for all modules individually. These are dominated by two effects: (1) track extrapolation uncertainties due to multiple scattering and (2) hit position uncertainties coming from the hit reconstruction algorithms. Both effects being random, they average out to a good approximation to zero if data from a sufficient number of hits is available. Misalignment is a systematic effect on these distributions. Therefore we determine the median (the 0.5~quantile of a distribution) for each module's distribution in order to measure such a systematic bias. These are then histogrammed for each subdetector, restricting to modules with at least 30~hits to ensure a large enough sample. Results are shown in figure \ref{fig.dmrplots} and in table \ref{tab.dmrresults}, compared with data from two Monte-Carlo studies where the tracker has been simulated assuming an ideal tracker geometry and after the alignment with data. Overall this shows that the alignment is already close to design specifications. Following the definitions of DMR, this is only an estimate of the modules' positions.

\begin{table*}
\caption{\textbf{Results from DMR plots.} RMS values of the distributions in the DMR plots (figure \ref{fig.dmrplots}) are given. Observe that this data covers the parts of the tracker hit by the cosmic ray particles. Especially in the pixel endcaps (PXE) the illumination is low due to the small size of the modules and the suboptimal track angles.  MC simulations were carried out using the misaligned and ideal geometry as starting point (column ``combined'' and ``ideal'' respectively).}
\label{tab.dmrresults}
{\footnotesize 
\begin{center}
\begin{tabular}{c|cccc|cc|c}
subdetector &   non-aligned &  global   & local    & combined & combined & ideal     & modules \cr
(coordinate)&    [$\mu$m]   &  [$\mu$m] & [$\mu$m] & [$\mu$m] & MC [$\mu$m] & MC [$\mu$m] & $>$30~hits \cr
	\hline
	PXB  ($u'$) &   329  &  7.5 &  3.0 & \textbf{ 2.6} &  2.1 &  2.1 & \cr
	PXB  ($v'$) &   274  &  6.9 & 13.4 & \textbf{ 4.0} &  2.5 &  2.4 & \raisebox{1.5ex}[-1.5ex]{757/768}\cr
	PXE  ($u'$) &   389  & 23.5 & 26.5 & \textbf{13.1} & 12.0 &  9.4 & \cr
	PXE  ($v'$) &   386  & 20.0 & 23.9 & \textbf{13.9} & 11.6 &  9.3 & \raisebox{1.5ex}[-1.5ex]{391/672}\cr
	TIB  ($u'$) &   712  &  4.9 &  7.1 & \textbf{ 2.5} &  1.2 &  1.1 & 2623/2724\cr
	TOB  ($u'$) &   169  &  5.7 &  3.5 & \textbf{ 2.6} &  1.4 &  1.1 & 5129/5208 \cr
	TID  ($u'$) &   295  &  7.0 &  6.9 & \textbf{ 3.3} &  2.4 &  1.6 & 807/816 \cr
	TEC  ($u'$) &   217  & 25.0 & 10.4 & \textbf{ 7.4} &  4.6 &  2.5 & 6318/6400 \cr
\end{tabular}
\end{center} 
}
\end{table*}

\subsection{Overlap studies}

\begin{figure}
	\begin{center}
	\includegraphics[width=7cm]{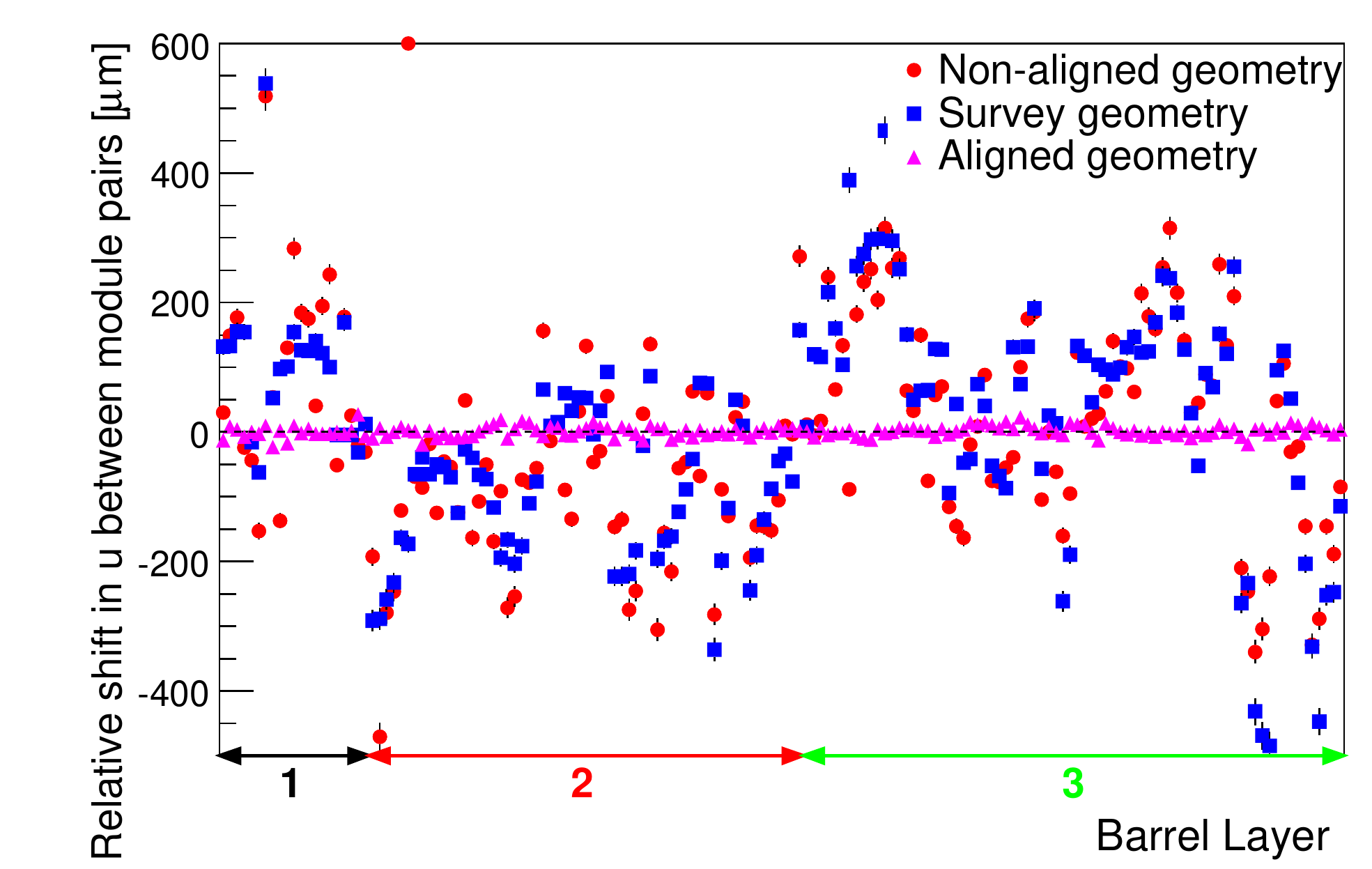}

	\smallskip

	\includegraphics[width=7cm]{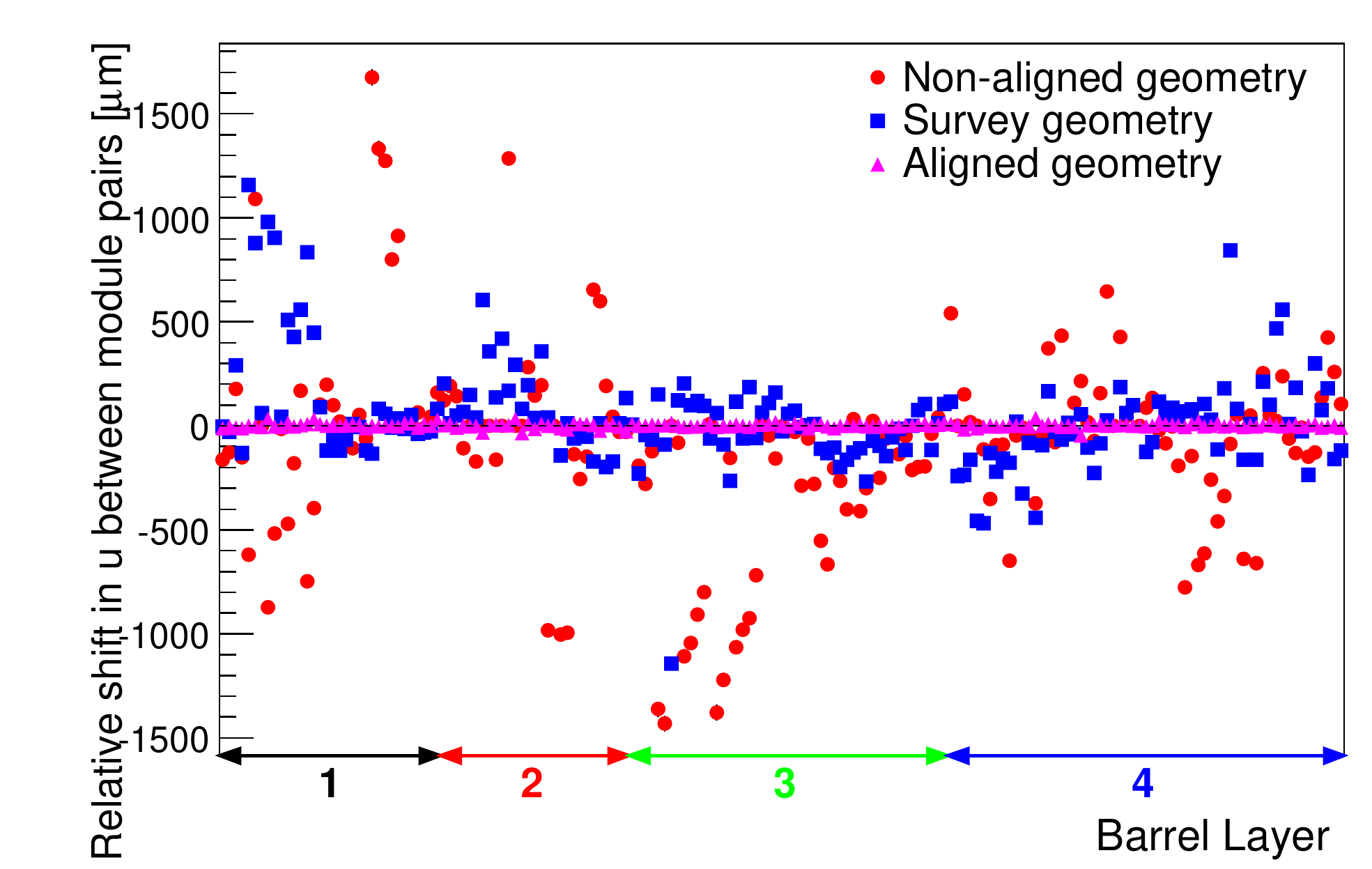}
	\end{center}
	\caption{\textbf{Overlap studies.} The upper plot shows results in PXB (local $u$), the lower one in TIB. The modules are plotted grouped by layer and show the relative shift in the local $u$ coordinate (in $r\phi$ direction, parallel to the most sensitive direction, i.e. perpendicular to the strip orientation) between overlapping module pairs. In the pixel barrel (PXB), the survey did not cover overlapping modules, therefore no improvement is visible for survey alone. In TIB, the plot clearly shows that survey improves the alignment. Nevertheless, the best results were obtained after the alignment has been carried out.}
	\label{fig.overlapplots}
\end{figure}

There are regions of the tracker where modules have overlap. This reduces the effects of multiple scattering due to geometric reasons. The residuals from the two neighbouring modules, obtained in the same manner as in the previous method, are compared. From the results plotted in Figure \ref{fig.overlapplots}) it can clearly be seen that the alignment performs well.

\subsection{Track parameter resolution}

\begin{figure}
	\begin{center}
	\includegraphics[width=8cm]{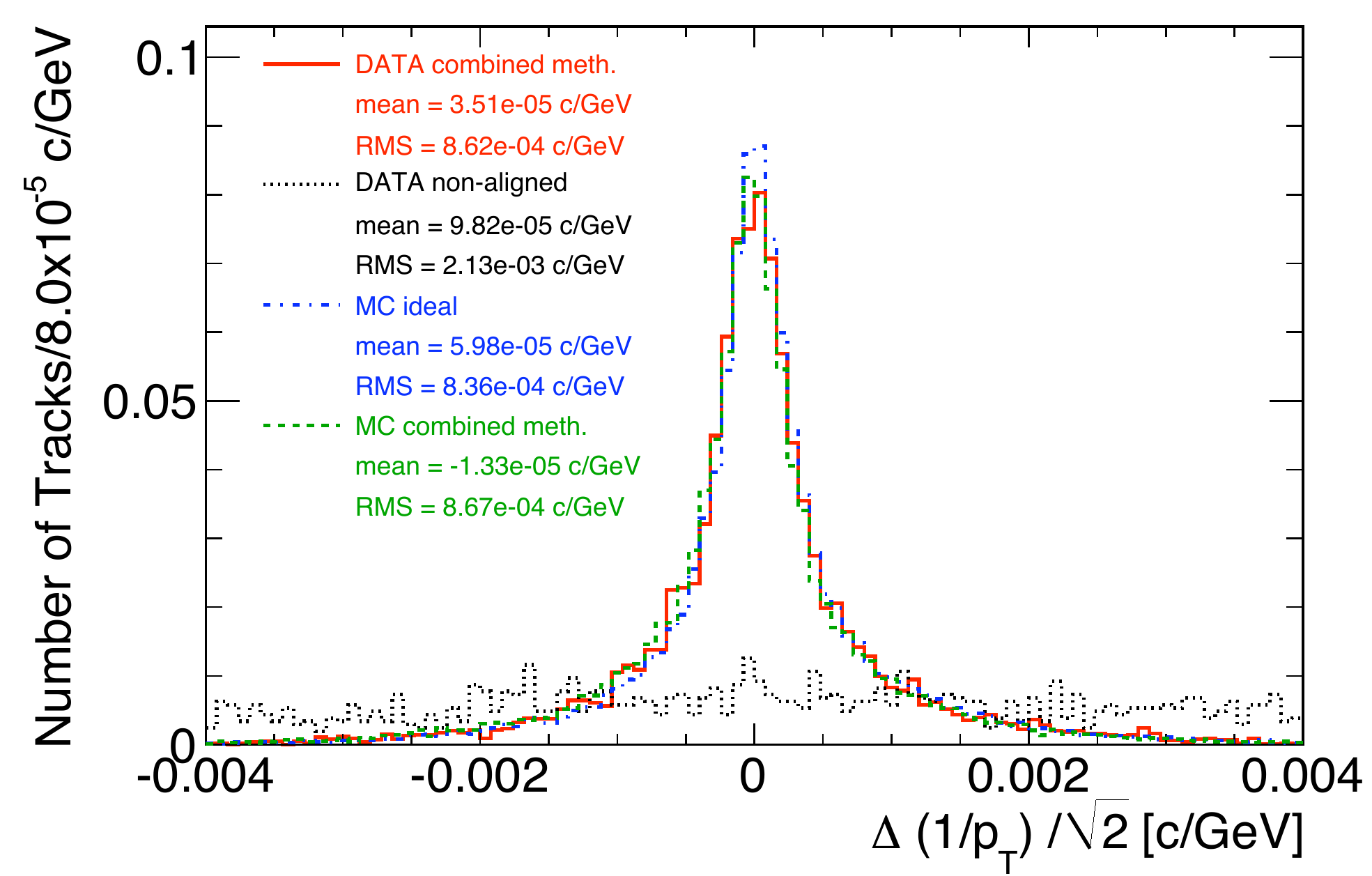}
	\includegraphics[width=8cm]{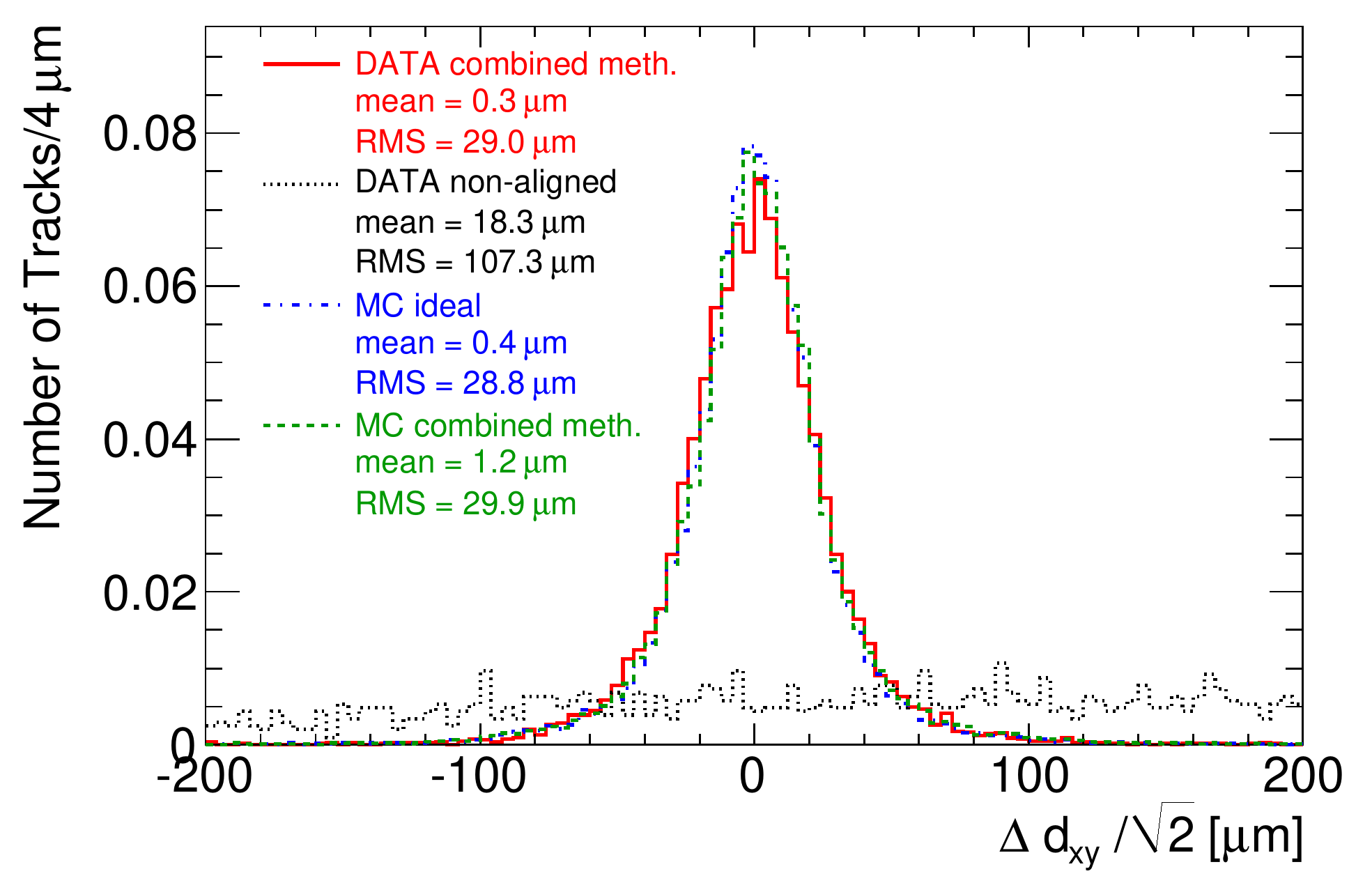}
	\includegraphics[width=8cm]{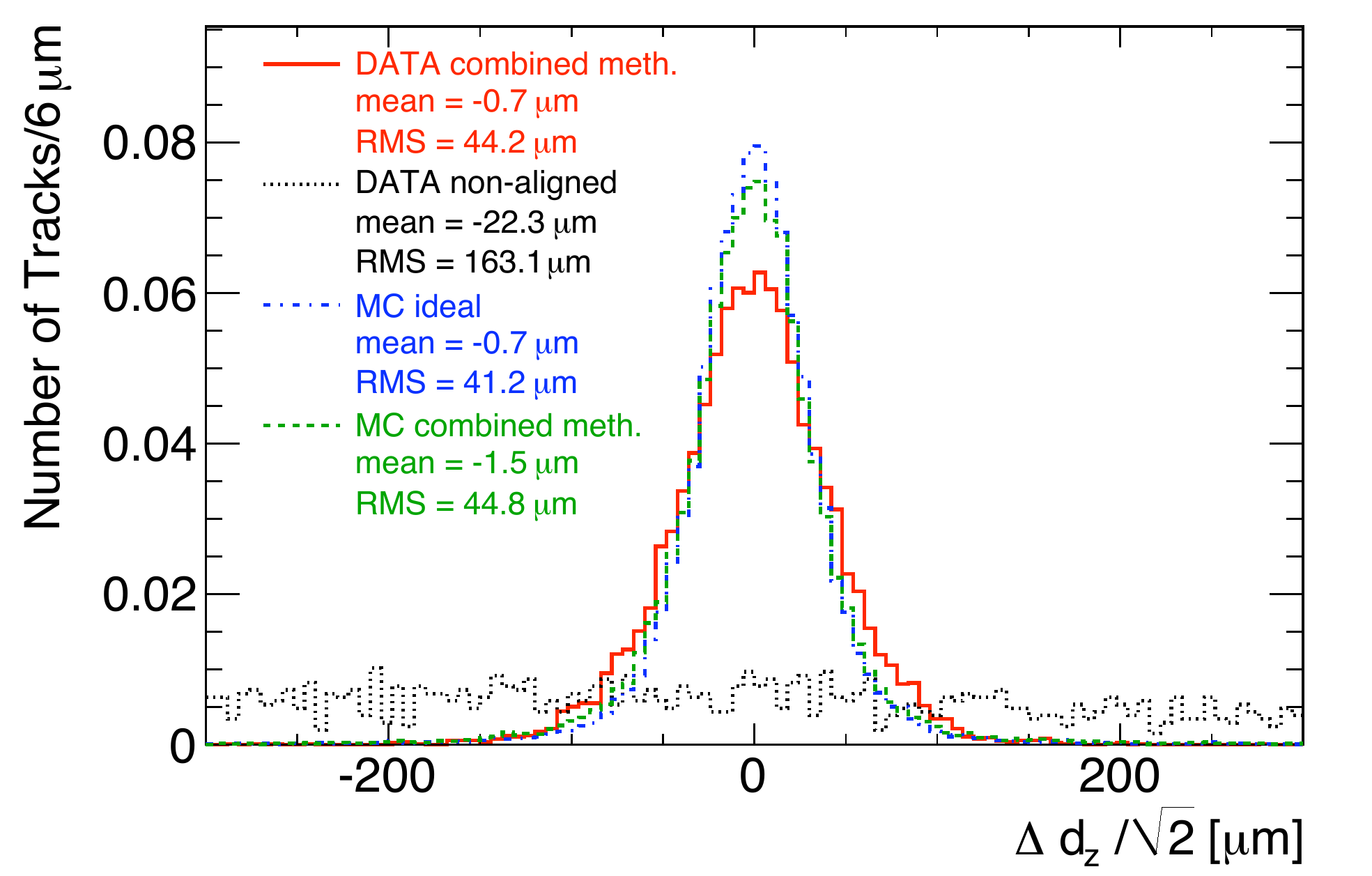}
	\end{center}
	\caption{\textbf{Resolution plots for $p_T$ and impact parameters.} The uppermost plot shows the distribution of the curvature $1/p_T$, the two following plots show the distribution of the impact parameter in the $xy$-plane and along $z$ respectively. All of them are compared to unaligned geometry and the results from a Monte-Carlo simulation. The aligned detector compares very well to the expected performance in Monte-Carlo (labelled as \emph{MC ideal}).}
	\label{fig.resoplots}
\end{figure}

The previously presented results are low-level measures of alignment performance. To get an impression on how the tracker operates under its intended use, tracks penetrating the pixel barrel have been selected. Such tracks were split at the closest approach to the geometric center of the tracker and refitted as separate tracks. Then the track parameters were compared at the closest approach of the two tracks. This procedure mimicks collision tracks as if they would originate from a common vertex within the pixel volume. Distribution plots for all track parameters show that the tracker indeed performs close to design specifications. Plots for the distribution of $p_T$ and for the impact parameters are shown in figure \ref{fig.resoplots}.

\section{Conclusions}

The studies presented here have shown that we are capable of aligning the inner tracker of CMS close to design specifications. No conclusion can be made for parts insufficiently illuminated by cosmic rays and remaining distortion modes leaving $\chi^2$ invariant. Using tracks from proton collisions will resolve this. We are looking forward to the start of data taking under beam conditions, where we will continue our efforts to align the inner tracker as closely to design specifications as possible.



\end{document}